\documentclass[nohyper,12pt,letterpaper,latexsym]{JHEP}
\usepackage{epsf}
\usepackage{epsfig}
\usepackage{cite}
\usepackage{amssymb,amsmath}

\title{ Non-minimally Coupled Tachyonic Inflation
in  Warped String Background}

\author{Pravabati Chingangbam, Sudhakar Panda\\
Harish-Chandra Reseach Institute\\
Chhatnag Road, Jhushi, Allahabad 211019, India\\
{\rm Email:} \email{prava@mri.ernet.in}, \email{panda@mri.ernet.in} }

\author{Atri Deshamukhya\\
Department of Physics, Gurucharan College\\
Silchar 788004, India \\
{\rm Email:} \email{atri\_d@yahoo.com}}

\abstract{ We show that the non-minimal coupling of tachyon field to the 
scalar
curvature, as proposed by Piao et al, with the chosen coupling parameter does
not produce the effective potential where the tachyon field can roll down 
from $T=0$ to large $T$ along the slope of the potential. 
We find a correct choice of the parameters which ensures this requirement 
and support slow-roll inflation. However, we find that the 
cosmological parameter found from the analysis of the theory are not in 
the range obtained from observations.  
We then invoke warped compactification and varying 
dilaton field over the compact manifold, as proposed by Raeymaekers, 
to show that in such a setup 
the observed parameter space can be ensured. }


\begin{document}


\section{Introduction}
Inflation \cite{guth} is possibly the only known mechanism which 
dynamically solves the flatness and the horizon problem of the universe. 
Thus it has become an almost indispensible ingredient in cosmology. The 
inflaton, a scalar field, can also produce the density perturbations 
causally which can match with the data from observation. For example, the 
recent WMAP data \cite{wmap} strongly supports the idea that the early 
universe went through an inflationary phase. Usually one considers the 
inflationary phase to be driven by the potential of a scalar field. 
Recently there has been an upsurge in activity for constructing such 
models in 
string theory, for example see \cite{quevedo} for a review. In the context 
of string theory, the tachyon field in the world volume theory of the open 
string stretched between a D-brane  and an anti-D-brane or on a non-BPS 
D-brane has been taken as a natural candidate to play the role of the 
inflaton \cite{cosmoref}. This possibility of the tachyon field driving 
the cosmological inflation is related to the decay of unstable brane as a 
time dependent process which was advocated by Sen \cite{sen}. The 
effective action used in the study of tachyon cosmology consists of the 
standard Einstein-Hilbert action and an effective action for the tachyon 
field on unstable D-brane or brane-antibrane system. What 
distinguishes the tachyon action from the standard Klein-Gordon form for 
scalar field is that the tachyon action, as we will see in the next 
section, is non-standard and is of the Dirac-Born-Infeld form \cite{dbi}. 
The tachyon potential is derived from string theory itself and has to 
satisfy some definite properties to describe tachyon condensation and  
other requirements in string theory. Thus it is an ideal situation to test 
if the tachyon field has any cosmological relevance. However, 
in \cite{koflin} it was shown that the slow-roll condition for the 
tachyon inflation is not possible to be achieved in conventional 
toroidal string compactification 
and within the validity of small coupling effective theory. To be precise, 
the slow-roll conditions were not compatible with the string coupling to 
be much less than one and the dimensionless parameter $v$, related to the 
volume of the compact space, to be much greater than one. 
This leads to the density fluctuations 
produced during the inflation being incompatible with COBE normalisation. 
The main source of this criticism stems from the string theory motivated 
values of the parameters in the tachyon potential i.e. the tension 
of the 
brane and the parameters in the gravitational coupling in four 
dimensions obtained via conventional toroidal string compactification. 
This 
objection has cast a shadow on the string motivation of this 
inflationary scenario. Nevertheless, as shown in \cite{steerver}, if one 
relaxes the string theory constraints on the above mentioned parameters 
i.e. if one takes a phenomenological approach, this theory naturally leads 
to inflation. More interestingly, these authors observed that the 
tachyon inflation does not lead to the same predictions as standard single 
field inflation, for example, there is a deviation in one of the second 
order consistency relations. They also noted that not only the tachyon 
inflation cannot be ruled out by current observations but also the planned 
observations probably cannot discriminate between tachyon inflation and 
single scalar field inflation. This may, however, change in future.

To circumvent the objection in \cite{koflin} within the conventional 
string compactification, Piao et al\cite{piao} 
introduced non-minimal coupling of tachyon field to gravity. The idea of 
non-minimal coupling of scalar fields with gravity has been implemented in 
the past for various applications \cite{maeda}. Piao et al claimed 
that for tachyon cosmology this can predict the observed density 
perturbation at the cost of the 
string scale being of the order of a few TeV.

In this note, we point out that the work of Piao et al 
\cite{piao} has a 
serious algebraic flaw  - rectification of which annuls all their 
predictions. In fact, the correction of this mistake, as we will see in 
section 2,  leads 
the effective potential being such that the tachyon field cannot roll down 
and hence the rest of their analysis becomes irrelevant. 
We reanalyze the 
inflationary scenario by choosing suitable non-minimal 
coupling parameters which modify the effective potential as required 
so that the tachyon field can evolve from its zero value to a non-zero 
value. Our motive is to see whether the new non-minimal 
coupling parameters can actually reproduce required values of   
cosmological observables. But the results are found to be negative, 
 contrary to the claim by the authors of \cite{piao}. It is 
found that the volume 
parameter $v$ of the compactified space remains to be  much less than one 
if it is 
chosen to fit the observed spectral index and density perturbations, 
contrary to the string theory requirement that $v \gg 1$ for the 
effective field theory of the tachyon-gravity system to be 
meaningful. We then 
redo the analysis in a warped compactification background following the 
analysis of \cite{joris} which considered the case of minimal coupling in 
such a warped background. Our analysis shows that we can have $v\gg 1$ for 
a wide range of string coupling constant $g$ and the minimum number of 
D6-branes required to produce the warped background is $10^{13}$ i.e., 
marginally less compared to the case in \cite{joris}.   

The paper is organized as follows:  
in section 2 we discuss the inflationary scenario of 
non-minimal coupling of tachyon field with gravity correcting the 
error alluded to earlier. In Section 3, we 
study this problem in a 
warped background and discuss the possibility of obtaining the 
cosmological parameters consistent with observations. Section 4 is 
devoted to discussion of our results.

\section{Non-minimally coupled tachyon-gravity }

We consider the following action for tachyon non-minimally coupled to 
gravity \cite{piao}
\begin{equation}
S = \int d^4x {\sqrt{-g}} \Big( {M_{P}^2\over{2}} R f(T) -
A V(T) {\sqrt{1+ B g^{\mu\nu} \partial_{\mu}T \partial_{\nu} T}} \Big)
\label{nonaction1}
\end{equation}
where $f(T)$ is a function of the tachyon $T$ and corresponds to the 
non-minimal coupling
factor. Note that for $f(T) = 1$ this action  
corresponds to that of the theory with minimal 
coupling. 
Here $V(T)$ is the positive definite tachyon potential which has a 
maximum at $T=0$ with normalization $V(0) = 1$ and $V \rightarrow 0$ as $T 
\rightarrow \infty$. $A$ and $B$ are 
dimensionful constants which depend on string length and the 
dilaton which sets the strength of the closed string coupling $g$.
In the conventional compactification (unwarped and constant 
dilaton)  
$A$, $B$ and $M_{P}^2$ are given by 
\begin{eqnarray}
A &=& {{\sqrt{2}}\over{(2\pi)^3 g \alpha'^2}} \label{A}\\
B &=& 8\ln 2 \alpha' \label{B} \\
M_{P}^2 &=& {v \over g^2\alpha'}
\end{eqnarray}
corresponding to the case of space-filling non-BPS D3-brane in type IIA 
theory. Here $v$ is a dimensionless constant 
related to the volume $V_6$ of the 
compactified manifold by
\begin{equation}
v = {2 V_6 \over (2\pi)^7 \alpha'^3}
\end{equation}
For the validity of the effective action we require $g < 1$ 
and $v \gg 1$.  The expressions for $A$, $B$ and $v$ will change 
for warped compactification as we will 
discuss later.

The action (\ref{nonaction1}) can be brought to the usual form of 
Einstein-Hilbert and 
matter action, for which it is simpler to derive the equation of motion, 
energy density and pressure,    
by performing a conformal transformation 
$ g_{\mu\nu}(x) \rightarrow f(T) g_{\mu\nu}(x) $, which yields 
\begin{equation}
S = \int d^4 x {\sqrt{- g}} \Big( {M_{P}^2 \over 2}  
             \big( R - {3 \over 2}
             {f'^2 \over f^2} g^{\mu\nu} 
             \partial_{\mu}T\partial_{\nu}T \big) 
             - A \tilde{V}(T)
             {\sqrt{1 + B f(T) g^{\mu\nu}
             \partial_{\mu}T\partial_{\nu}T}} \Big) \label{nonaction2} 
\end{equation}
where $\tilde{V}(T)=V(T)/f^2$ is now the effective potential of the 
tachyon. The energy density and pressure are found to be  
\begin{eqnarray}
\rho &=& {A \tilde{V} \over \sqrt{ 1- B f \dot{T}^2} }
        + {3 \over 4} M_{P}^2 {f'^2 \over f^2} \dot{T}^2 \\
p &=&  - A \tilde{V}\sqrt{1 - B f \dot{T}^2}
         +{3 \over 4} M_{P}^2 {f'^2 \over f^2} \dot{T}^2
\end{eqnarray}

The equation of motion  of the tachyon and the Friedmann equation are 
$$ \ddot{T} \left[ {1 \over {1 - B f \dot{T}^2}}
+ {3 \over 2} M_{P}^2 {f'^2 \over f^2}
{\sqrt{1 - B f \dot{T}^2} \over A B f \tilde{V}} \right]
+ 3 H \dot{T} \left[ 1 + {3 \over 2} M_{P}^2 {f'^2 \over f^2}
{\sqrt{1 - B f \dot{T}^2} \over A B f \tilde{V}} \right]
$$
\begin{equation}
 + \dot{T}^2 {f' \over 2 f} \left[
{1 \over {1 - B f \dot{T}^2}} + 3 M_{P}^2
\left( {{ff''-f'^2} \over f^2} \right)
{\sqrt{1 - B f\dot{T}^2} \over A B f\tilde{V}} \right]
+{\tilde{V}' \over B f\tilde{V}} = 0 \label{noneom}
\end{equation}
\begin{equation}
H^2 = {1 \over 3 M_{P}^2} 
      {A \tilde{V} \over \sqrt{1 - A f \dot{T}^2}}
      + {1 \over 4} {f'^2 \over f^2} \dot{T}^2 \label{nonfried}
\end{equation}
The inflationary condition is obtained to be

\begin{equation}
B f \dot{T}^2 \left( 1 + {M_{P}^2 \over A B f \tilde{V}} {f'^2 \over 
f^2}
\sqrt{1 - B f \dot{T}^2} \right) < {2 \over 3}
\end{equation}
With slow-roll approximation, eqns. (\ref{noneom}) and 
(\ref{nonfried}) can be rewritten as   
\begin{eqnarray}
3 H \dot{T} + {\tilde{V}' \over B f\tilde{V}} 
             \cong 0 \label{nonsloweom}\\
H^2 \cong {1 \over 3 M_{P}^2} A \tilde{V}
\end{eqnarray}
where we have assumed that
\begin{equation}
\delta \equiv {3 M_{P}^2 \over 2}
              {f'^2 \over f^2} {1 \over A B f\tilde{V}} \ll 1
\end{equation}
and which, as we will see, is justified for an appropriate choice of
$f(T)$. 
The slow-roll parameters are found to be  
\begin{eqnarray}
\epsilon_1 &=& {3 \over 2} B f \dot{T}^2 
               = {M_{P}^2 \over 2 ABf} 
                  {\tilde{V}'^2 \over \tilde{V}^3}  \\
\epsilon_2 &=& {2 \ddot{T} \over H \dot{T}}
               = {M_{P}^2 \over ABf} \left( 
                 {f' \tilde{V}' \over f \tilde{V}^2}
                 + 3 {\tilde{V}'^2 \over \tilde{V}^3}
                 - 2 {\tilde{V}'' \over \tilde{V}^2} \right)
\end{eqnarray}
and
\begin{eqnarray}
\epsilon_2 \epsilon_3 
        &=& {M_{P}^2 \over (AB)^2} \left(
            { \dddot{T} \over H \dot{T} } 
           - { \ddot{T} \over  \dot{T} }{ \dot{H} \over H^2 }
           - { \ddot{T}^2 \over H \dot{T}^2 } \right) \nonumber\\
{}      &=& {M_{P}^2 \over (AB)^2} \Big(
            {2 \tilde{V}'\tilde{V}'' \over f^2 \tilde{V}^4}
            - 10 {\tilde{V}'^2 \tilde{V}'' \over f^2 \tilde{V}^5}
            + 9 {\tilde{V}'^4 \over f^2 \tilde{V}^6}
            - { f''\tilde{V}'^2 \over f^3 \tilde{V}^3} \nonumber \\
{}      &{}& - {f' \tilde{V}'\tilde{V}'' \over f^3 \tilde{V}^4}
            + 5 {f'\tilde{V}'^3 \over f^3 \tilde{V}^5}
            + 2 {f'^2 \tilde{V}'^2 \over f^4 \tilde{V}^4} \Big)
\end{eqnarray}
The usual slow-roll parameters denoted by $\epsilon$ and $\eta$ are 
related to $\epsilon_1$ and $\epsilon_2$ by $\epsilon=\epsilon_1$ and 
$\eta = 2 \epsilon_1 - {1\over 2} \epsilon_2$. Slow-roll conditions are 
$\epsilon_1 \ll 1$ and $|\epsilon_2| \ll 1$. The end of inflation is given 
by $|\epsilon_2(T_e)| \simeq 1$, where $T_e$ is the value of the field at  
the end of inflation.

The number of e-folds 
between an arbitrary initial value of the field $T$ and $T_e$ is given by 
\begin{equation}
N_e (T) \simeq {AB \over M_{P}^2} \int^{T}_{T_e} {\tilde{V}^2 \over 
\tilde{V}'} dT
\end{equation}

To first order in the slow-roll parameters, the scalar and 
gravitational power spectra are given by
\begin{eqnarray}
\mathcal{P}_{\mathcal{R}} (k) &=& {H^2 \over 8 \pi^2 M_{P}^2 
\epsilon_1} = {1 \over 24 \pi^2} {A\over M_{P}^4}
                {\tilde{V}\over \epsilon_1}   \label{scalar}  \\
\mathcal{P}_g (k) &=& {2 H^2 \over \pi^2 M_{P}^2} \label{tensor}.
\end{eqnarray}
In the above equations the right hand side is to be evaluated at 
$a H = k$ where $a$ is the scale factor, $H$ being the Hubble parameter. 
The tensor-scalar ratio $r$, the scalar spectral index $n$
and the tensor spectral index $n_T$ are given by:
\begin{eqnarray}
r &\equiv& {\mathcal{P}_g \over \mathcal{P}_{\mathcal{R}} } 
                    = 16 \epsilon_1 \label{r}\\
n - 1 &\equiv& { d \ln \mathcal{P}_{\mathcal{R}} (k) \over d \ln k} 
               = - 2 \epsilon_1 - \epsilon_2
               = -6\epsilon + 2\eta \\
n_T &\equiv& { d \ln {\mathcal{P}}_g (k) \over d \ln k} 
             = - 2 \epsilon_1.
\end{eqnarray}

The running of the spectral indices is 
\begin{eqnarray}
{d n \over d \ln k} &=& -2 \epsilon_1 \epsilon_2 - \epsilon_2 
\epsilon_3 \label{dndlnk}\\
{d n_T \over d \ln k} &=& -2 \epsilon_1 \epsilon_2
\end{eqnarray}

The Gaussian potential $V(T)=e^{-T^2}$ (motivated from boundary string 
field theory), gives a good description for small $T$ and can be assumed 
to be accurate for the inflationary epoch.  
With this the effective potential becomes $\tilde{V} = e^{-T^2}/f^2$. 
In general $f(T)$ can be expanded as $f(T) = 1 + \sum_{i=1} c_i T^{2i}$. 
Expanding $\tilde{V}$ in powers of $T$ we get  
\begin{equation}
\tilde{V}(T) = 1-(1+2c_1)T^2 + ({1\over 2} + 2c_1 - 2c_2
                           +3 c_1^2)T^4 + \ldots \label{pot}
\end{equation}
With this effective potential, the requirement that 
in the neighbourhood of $T=0$ we should have the slow-roll condition 
$|\eta| \ll 1$ gives the constraint:
\begin{equation}
{4 M_{P}^2 \over  AB} (1 + 2 c_1) \ll 1
\end{equation}
Note that for  $c_1=0$ this gives the unjustifiable slow-roll constraint  
$g/v >> 127$ 
\cite{koflin} in the minimal coupling case. 
However, the choice $c_1 = -1/2$ makes the slow-roll 
condition eqn.(27) automatically  satisfied, as noted in \cite{piao}. 
Further, if all other $c_i$'s are chosen to be zero, the expansion of 
$\tilde{V}$ become
\begin{equation}
\tilde{V}(T) = 1 + {1\over 4}T^4 + \ldots 
\end{equation}
This potential has a minimum at $T=0$ and it increases as $T$ increases, 
Thus, it is not suitable to have a 
slow-roll inflationary scenario with this potential, where the inflaton 
field rolls down from a lower value to higher value.

Piao et al have made the following error. They had the following expansion 
of $\tilde{V}$
\begin{equation}
\tilde{V}(T) = 1-(1+2c_1)T^2 + ({1\over 2} + 2c_1 - 2c_2
                           - c_1^2)T^4 + \ldots 
\end{equation}
with the incorrect term $-c_1^2$ instead of $3 c_1^2$ in the 
coeffiecient of $T^4$, as in eqn.(26). With the choice $c_1 = -1/2$ and 
all other $c_i$'s 
put equal to zero, the above expression then is 
$$ \tilde{V}(T) = 1 - {3\over 4} T^4 + \ldots $$ 
with the potential having a maximum at $T=0$. 
Thus all their remaining analysis based on this incorrect potential breaks 
down. Further, they claimed that comparing with observational data on 
$\mathcal{P}_R$ and $\mathcal{P}_g$ it is possible 
to achieve $g<1$ and $v\gg 1$. However, when observational data on spectral 
index etc are taken into account this claim is also found to be incorrect. 

It is, nevertheless, possible to have 
a feasible model of slow-roll inflation, with the potential having a 
maximum at $T=0$, by invoking non-minimal
coupling of tachyon with gravity with the choice $c_1 = -1/2$ (so 
that $T^2$ 
term vanishes) and keep at least upto $T^4$
term in $f(T)$, choosing $c_2$ in such a way that the coefficient of $T^4$ in the 
expansion of $\tilde{V}$ is negative. We choose, accordingly, $c_2$ to be 1/4 
and all other $c_i$'s to be zero. In this case we have
$$ \tilde{V}(T) = 1 - {1\over 4} T^4 + \ldots $$
This effective potential has maximum at $T=0$ as required.
For our above choice of $c_1$ and $c_2$, we find that $\delta$  
is of the same order as  
$\epsilon_2$ (or $\eta$). Thus the requirement of $\delta \ll 1$ 
is automatically satisfied within the slow-roll regime
and is not an independent requirement.  

Following \cite{steerver} we find that $\tilde{V}'' < 
\tilde{V}'^2/\tilde{V}$ for all values of $T$.
Inflation can take place in either region I, given by $\tilde{V}''\le 0$,
or region II, given by  $0 \le \tilde{V}'' \le \tilde{V}'^2/\tilde{V}$. 
Here region I ends roughly at $T=1.12$. This means that $T_e$ can at most
have this value and correspondingly $T_*$, which is the value of the 
tachyon field at roughly 60 e-folds when the cosmological scale crosses 
the horizon, is bounded 
to a maximum value of roughly 0.5. We restrict ourselves to region I. 
All observables, namely, $n$, $\mathcal{P}_R$, $r$ and $dn/d{\rm ln} k$ 
are to be evaluated at $T_*$. 

Clearly the advantage of introducing non-minimal coupling with the choice
of parameters $c_1=-1/2$ and $c_2 = 1/4$ is to remove the slow-roll
constraint that $AB / 4 M_{P}^2 \gg 1$ and hence it is possible to choose
it to be small. It is not arbitrary, however, and must be chosen to fit
observations. We outline below the strategy of our numerical estimation:
\begin{enumerate} 
\item From the condition for the end of inflation,
$\epsilon_2(T_e) = 1 $, $T_e$ is obtained as a function of $AB/M_{P}^2$.

\item Fixing $N_e$ to be 60 we evaluate $T_*$ as a function of $T_e$ and
consequently a function of $AB/M_{P}^2$.  

\item Then we find the range of $AB/M_{P}^2$ for which the spectral index 
is in the range $0.94\le n\le 1$.   

\item $\mathcal{P}_R$ independently depends on $A/M_{P}^4$ and its range 
cannot be fixed only from $AB/M_{P}^2$. 
We use the observational input $\mathcal{P}_R \le 0.71 \times 2.9 \times 
10^{-9}$ \cite{peiris} in eqn.(19) to get the range for $A/M_{P}^4$.

\end{enumerate}

We quote below the range of values for ${AB/M_{P}^2}$, $A/M_{P}^4$ and 
the observables $r$ and $dn/d{\rm ln} k$ obtained from our numerical 
estimates. 
The upper bound for ${AB/M_{P}^2}$ is obtained from the maximum value of
$T_e$ in region I. This constrains $AB/M_{P}^2$ as
\begin{equation}
10^{-13} \le {AB\over{M_{P}^2}} \le 69 \label{AB}
\end{equation}

The range for ${A/{M_{P}^4}}$ is:
\begin{equation}
1.7 \times 10^{-39} < {A\over{M_{P}^4}} < 9.74 \times 10^{-11}  
\end{equation}

The range for $r$ is obtained to be
$$5.6\times  10^{-32} < r < 2.5 \times 10^{-3} $$

The running of the spectral index is negative and is found to lie in  
the range
$$ -8.9\times 10^{-4} < {{dn}\over{d\ln k}} < -5.09 \times
10^{-4} .$$
$r$ and $ {{dn}/{d\ln k}}$ are well within the experimental 
bounds.

An analysis of the above estimates in terms of the string theory
parameters $g$ and $v$ is in order. For each corresponding set of values 
of 
$AB/M_P^2$ and $A/M_P^4$ we can solve for $g$ and $v$ as  
\begin{eqnarray}
g &=& {(A/M_P^4) \over (AB/M_P^2)^2} \times 0.175 \\
v &=& {(A/M_P^4) \over (AB/M_P^2)^3} \times 0.0055
\end{eqnarray}
We tabulate the corresponding 
values below for values of $n$  decreasing from 0.963 to 0.94 :

\vskip .5cm
\begin{center}
\begin{tabular}{|llll|}
\hline
$AB/M_P^2$ & $A/M_P^4$ & $g$ & $v $ \\
\hline
60 & $ 9.74\times 10^{-11}$ & $4.73\times 10^{-15}$ & $2.48 \times
10^{-18}$ \\
\hline
1 & $1.08 \times 10^{-13}$ & $1.89\times 10^{-14}$ & $5.94\times 
10^{-16}$
\\
\hline
$0.01$ & $1.21\times 10^{-17}$ & $2.12\times 10^{-14}$ & $6.68
\times 10^{-14}$ \\
\hline
$10^{-4}$ & $1.22\times 10^{-21}$ & $2.13\times 10^{-14}$ & 
$6.71\times 10^{-12}$ \\
\hline
$10^{-6}$ & $1.22\times 10^{-25}$ & $2.13\times 10^{-14}$ & 
$6.71\times 10^{-10}$ \\
\hline
$10^{-8}$ & $1.22\times 10^{-29}$ & $2.13\times 10^{-14}$ & $ 
6.71\times 10^{-8}$ \\
\hline
$10^{-10}$ & $1.22\times 10^{-33}$ & $2.13\times 10^{-14}$ & $6.71\times 
10^{-6}$ \\ 
\hline
$10^{-13}$ & $1.7 \times 10^{-39}$ & $2.96\times 10^{-14}$ & 
$1.93\times 10^{-3}$ \\
\hline
\end{tabular}
\vskip .5cm
Table 1. 
\end{center}

It is clear from the above table that, though $g$ is always much less than 
one, it is impossible to constrain $v$ to be much greater than one and 
still satisfy observation. 
Thus we conclude that even if we invoke non-minimal coupling 
with fine tuned coupling parameters $c_1$ and $c_2$, we cannot 
achieve $g<1$ and $v\gg 1$, satisfying the observational 
constraints on cosmological parameters.

\section{\bf{Tachyon 
Inflation in Warped Background}}

Since within conventional compactification it is not possible to obtain 
physical parameters of inflation consistent with observation, keeping 
$g<1$ and $v\gg 1$ even with non-minimal coupling, 
it becomes important to look for alternative string 
compactifications with a hope that this may improve  
the values of the parameters and allow us to have $g<1$ and $v\gg 1$,  
thus freeing tachyon inflation from the criticism
mentioned earlier.  Already some progress has been achieved in this
direction, recently, by Raeymakers \cite{joris}. The author considered a
warped compactification, where the four dimensional metric contains an
overall factor which can vary over the compactified space \cite{warp}. 
The dilaton was
allowed to vary over the compact manifold as well. The latter has two
advantages. Since the parameters governing the tachyon action depends upon
dilaton, one has more freedom by allowing it to vary. Secondly, the 
varying dilaton contributes to the warp factor in the Einstein frame. Such 
a warped background could be obtained by wrapping a large number of 
space-filling D-branes on a cycle of the compact manifold since the back 
reaction produces a ``throat region"  with enough warping and varying 
dilaton. It is
found that the parameter range required for inflation can be accommodated
in the background of D6-branes wrapping a three-cycle in type IIA theory.
The important ingredient which goes in here to make the inflation possible
is the fact that the string coupling decreases faster than the (string
frame) warp factor as one approaches the branes. 

The effective field theory that describes 
inflation when the tachyon is coupled minimally to gravity \cite{dbi}  
is the same as eqn.(1) when we set $f(T)=1$, i.e., 
\begin{equation} 
S = \int d^4x {\sqrt{-g}} \Big( {M_{P}^2\over{2}} R -  
A V(T) {\sqrt{1+ B g^{\mu\nu} \partial_{\mu}T \partial_{\nu} T}} \Big)
\label{action}
\end{equation}

Warping is introduced by 
considering the ten dimensional string frame metric of the form
\begin{equation}
ds^2 =  e^{2 C(y)} g_{\mu\nu}(x) dx^\mu dx^\nu + g_{mn}(y) dy^m dy^n
\label{metric}
\end{equation}
where $e^{2C}$ is the warp factor and can take very small values. 
Moreover, if the dilaton is allowed to vary over the compact manifold as 
\begin{equation}
\phi = \phi_0 + \phi(y),
\end{equation}
one finds that 
the four dimentional Planck mass in such warped compatification is
\begin{equation}
M_{P}^2 = {\tilde{v} \over g^2 \alpha' }.
\end{equation}
Here $g = e^{\phi_0}$ and the 'warped volume' $\tilde{v}$ is
\begin{equation}
\tilde v = {2 \over (2 \pi)^7 \alpha'^3} \int d^6 y \sqrt{g_6} e^{-2 \phi 
+ 2 C}.
\end{equation}
$\tilde{v}$ can be taken to be the same order as $v$ by choosing the 
average value of 
$e^{-2\phi+2C}$ to be of order one as taken in \cite{joris}. However,  
locally the $y$-dependent functions affect $A$ and $B$ in 
eqn.(\ref{action}). For example, when the non-BPS 
D3-brane is  embedded in such a  
background, it can be seen that $A$ and $B$ become   
\begin{eqnarray}
A &=& { \sqrt{2}  e^{4 C - \phi} \over (2 \pi)^3 g \alpha '^2}\label{A2} 
\\
B &=& 8 \ln 2 \alpha ' e^{-2C}.\label{B2}
\end{eqnarray}
Note that the functions $C$ and $\phi$ are not arbitrary but are subject 
to the solutions of equations of motion derived from supergravity theory.  
The warping modifies the slow-roll constraint $AB/M_P^2 \gg 1$  as  
\begin{equation}
{g \over v} e^{2 C - \phi} \gg 127.
\label{gvcond2}
\end{equation}
which can be satisfied in this background where locally we have 
\begin{equation}
 e^{2 C - \phi} \gg 1.
\label{gvcond3}
\end{equation}
This condition can be achieved by wrapping a large number $N$ of 
D6-branes over a three cycle.

Within the regime of supergravity approximation one can derive the 
following expressions for the warp factor and the dilaton: 
\begin{eqnarray}
e^{2C} &=& (g N_{\rm{min}})^{-2/3}  \label{C}\\
e^{-\phi} &=& g N_{\rm{min}} \label{phi}
\end{eqnarray}
where $N_{\rm{min}}$ is the minimum number of branes. In this case the 
condition (\ref{gvcond2}) becomes 
$$
{g^{4/3} \over v} \gg {127 \over N_{\rm min}^{1/3}}.
$$
Thus for satisfying the slow-roll condition, $N_{\rm min}$ is required to 
be at least of the order of $10^6$ \cite{joris}.   
Further, the author found that in order to satisfy the observational 
constraints on density perturbation and spectral index,  
the minimum number of branes required 
is of the order of $10^{14}$. 

To highlight the allowed values of  $g$ and $v$ we give below their 
relationship, along with the warp factors, with the parametric values of 
$A$ and $B$ obtained from observational data:
\begin{eqnarray}
ge^{\phi} &=& {(A/M_P^4) \over (AB/M_P^2)^2} \times 0.175 \\
v e^{-2C + 2\phi} &=& {(A/M_P^4) \over (AB/M_P^2)^3} \times 0.0055
\end{eqnarray}

Unlike the unwarped case we now have an extra  
parameter due to warping, which can be used to obtain the required values 
of $g$ and $v$. Eliminating $e^{\phi}$ and $e^{2C}$ using eqns.(\ref{C}) and 
(\ref{phi}) in the above equations 
we get a relation between $v$ and $g$: 
\begin{equation}
v = g^{4/3} y^{-1/3} x^{-1/3} \times 0.057
\end{equation}
where $x$ denotes $AB/M_P^2$ and $y$ denotes $A/M_P^4$. 
To have $v\gg 1$ the condition on $g$ is
\begin{equation}
g > y^{1/4} x^{1/4} \times 8.62
\end{equation}

The range of values obtained are tabulated below. Each row  
of  Table 2 corresponds to a different value of $n$, which    
increases from 0.94 to 0.97. The sixth column gives the 
constraint on $g$ when we demand $v\gg 1$:

\vskip .5cm
\begin{center}
\begin{tabular}{|llllll|}
\hline
$AB/M_P^2$ & $A/M_P^4$ & $n$ & $ ge^{\phi}$ & $ve^{-2C+2\phi}$ & $g$ \\
\hline
70.9 & $ 3.02\times 10^{-10}$ & $0.94$ & $1.05 \times 10^{-14}$ 
           & $4.66\times 10^{-18} $ & $g>0.10$ \\
100 & $ 6.54 \times 10^{-10}$ & $1.0\times 0.953$ & $10^{-14}$ & $10^{-18}$ & 
$g>0.14$\\
\hline
$500$ & $3.69 \times 10^{-9}$ & 0.969 & $2.58\times 10^{-15}$ & 
           $8.12 \times 10^{-19}$ & $g>0.32$ \\
\hline

$900$ & $5.92 \times 10^{-9}$ & $0.9696 $  & $1.28\times 10^{-15}$ 
       & $4.47 \times 10^{-20}$ & $g>0.41$ \\
\hline
$ 1100$  & $ 6.93\times 10 ^{-9}$ & $0.9697$ & $1.0\times 10^{-16}$ & 
       $2.86 \times 10^{-20}$  & $g>0.45$ \\
\hline
\end{tabular} 
\vskip .5cm
Table 2.
\end{center}

We see from the Table 2 that it is possible to have $v\gg 1$, but only for  
$g > 0.1$. 

We now repeat this analysis, i.e., to include warping for the non-minimal 
coupling case discussed in section 2. This is in expectation that 
since the non-minimal coupling removes the slow-roll constraint 
and warping improves the value of $v$, the parameters and the minimum 
number of branes can be improved. 

The numerical values we get are tabulated below. This is essentially the 
same as Table 1, but with the values of $g$ and $v$ reinterpreted in the 
light of the warped background, as presented in Table 2. 

\vskip .5cm
\begin{center}
\begin{tabular}{|llllll|}
\hline
$AB/M_P^2$ & $A/M_P^4$ & $ge^{\phi}$ & 
$v e^{-2C+2\phi}$ & $g$ & $ v$ for $g=0.1$\\
\hline
60 & $ 9.74\times 10^{-11}$ & $4.73\times 10^{-15}$ & $2.48 \times 
10^{-18}$ & $ g> 0.07$ & $1.47$\\
\hline
1 & $1.08 \times 10^{-13}$ & $1.89\times 10^{-14}$ & $5.94\times 10^{-16}$ 
& $g> 0.0049$ & $55.56$\\
\hline
$0.01$ & $1.21\times 10^{-17}$ & $2.12\times 10^{-14}$ & $6.68 
\times 10^{-14}$ & $ g>1.61 \times 10^{-4}$ & $5.35\times 10^{3}$\\
\hline
$10^{-4}$ & $1.22\times 10^{-21}$ & $2.12\times 10^{-14}$ & 
$6.71\times 10^{-12}$ & 
$g>5.09 \times 10^{-6}$ & $ 5.33\times 10^{5}$\\
\hline
$10^{-6}$ & $1.22\times 10^{-25}$ & $2.12\times 10^{-14}$ & $6.71\times  
10^{-10}$ & 
$g>1.61\times 10^{-8}$ & $5.33 \times 10^{7}$ \\
\hline
$10^{-8}$ & $1.22\times 10^{-29}$ & $2.13\times 10^{-14}$ & 
$6.71\times 10^{-8}$ & $g>5.09 \times 10^{-9}$  & $5.33 \times 10^{9}$\\
\hline
$10^{-10}$ & $1.22\times 10^{-33}$ & $2.13\times 10^{-14}$ & 
$6.71\times 10^{-6}$ & $g>1.61\times 10^{-10}$  & $5.33\times 
10^{11}$\\
\hline
$10^{-13}$ & $1.7\times 10^{-39}$ & $2.98\times 10^{-14}$ & 
$1.93\times 10^{-3}$ & $g>9.85\times 10^{-13}$  & $4.78\times 10^{14}$\\
\hline
\end{tabular}
\vskip .5cm
Table 3.
\end{center}

The last column of Table 3 quotes the values of $v$
for $g=0.1$ to highlight the improved values of $v$.
Thus, $v \gg 1$  is achieved over a wide range of values 
of $g$. The minimum number of 
branes required in this case turns out to be of the order of $10^{13}$, 
which is a marginal decrease of one order of magnitude from the minimal 
case.

We can also estimate, for this model, the scale of inflation which is 
found to range from $10^{16}$ to $10^{9}$ GeV. This is in contrast to the 
range from $9.8\times 10^{16}$ to $1.7\times 10^{16}$ GeV in the case of 
minimal coupling in a warped background \cite{joris}.

\section{Discussion}

In this paper, we have shown that the non-minimal coupling of 
the form $-RT^2/2$ between tachyon field 
and gravity, proposed in \cite{piao}, is not consistent with slow-roll 
inflation in cosmology. We correct this problem by adding another non-minimal 
coupling of the form $RT^4/4$. However, we found that even if slow-roll 
conditions are 
satistied, it does not meet the fundamental requirement that one 
needs $g<1$ 
and $v\gg 1$ for the validity of low energy effective theory. Thus, 
non-minimal coupling alone, as analyzed in section 2, does not circumvent the 
problems in tachyon cosmology. 
We then invoke 
the warped compactification with a varying dilaton field in the compact space 
and analyze this problem, closely following \cite{joris}. We find that this 
model can be a viable model for inflation in cosmology. However, we still 
need a large number of D6-branes ($10^{13}$, compared to $10^{14}$ in the 
minimal case) for producing the required warped background. We note that 
these non-minimal coupling terms in the effective 
theory can arise by computing the S-matrix elements of open string 
tachyons  
and closed string gravitons \cite{coupling}. Though we kept only quadratic 
and 
quartic terms in tachyon field, in principle we should consider the coupling 
of tachyons to all orders to obtain an exact form involving non-minimal 
couplings. We will return to this issue in future. 
It is possible that the exact non-minimal coupling to 
all orders in tachyon field can produce an 
effective potential which can have a local minimum. Such a minimum can help 
in addressing the reheating issue, just as the case of any other scalar field 
driven cosmology having a potential with a true minimum. This possibly will 
be a suitable alternative to the proposal of \cite{cline} in the context of 
tachyon inflation. We also point out that our work assumes the possibility of 
stabilizing various scalar moduli fields and other requirements of warped 
compactification \cite{kklt} just as in \cite{joris}.

\noindent{\bf{Acknowledgment}}

We thank Ashoke Sen and Sandip Trivedi for many useful discussions.

\end{document}